\documentclass[conference]{HPCStrans}
\usepackage{graphicx}
\usepackage{amsmath}
\usepackage{url}
\usepackage{multirow}
\usepackage{pbox}

\usepackage{txfonts}

\usepackage{caption}
\usepackage{array}

\usepackage[caption=false,font=footnotesize]{subfig}
\usepackage{tikz}
\usepackage{textcomp}
\usepackage{hyperref}
\usepackage{lipsum}

\newcommand\copyrighttext{%
  \footnotesize \textcopyright 2016 IEEE. Personal use of this material is permitted.
  Permission from IEEE must be obtained for all other uses, in any current or future
  media, including reprinting/republishing this material for advertising or promotional
  purposes, creating new collective works, for resale or redistribution to servers or
  lists, or reuse of any copyrighted component of this work in other works.
  DOI: \href{http://ieeexplore.ieee.org/document/7568421/}{10.1109/HPCSim.2016.7568421}}
\newcommand\copyrightnotice{%
\begin{tikzpicture}[remember picture,overlay]
\node[anchor=south,yshift=10pt] at (current page.south) {\fbox{\parbox{\dimexpr\textwidth-\fboxsep-\fboxrule\relax}{\copyrighttext}}};
\end{tikzpicture}%
}

\begin{document}
\title{BSMBench: A Flexible and Scalable HPC Benchmark from Beyond the Standard Model Physics}

\author{\IEEEauthorblockN{Ed Bennett, Biagio Lucini}
\IEEEauthorblockA{College of Science\\Swansea University\\Swansea, UK\\\{E.J.Bennett, B.Lucini\}@swansea.ac.uk}
\\
\IEEEauthorblockN{Kirk Jordan}
\IEEEauthorblockA{Thomas J Watson Research Center\\IBM Research\\Cambridge, MA, USA\\kjordan@us.ibm.com}
\\
\IEEEauthorblockN{Claudio Pica}
\IEEEauthorblockA{CP$^3$-Origins \& the Danish IAS\\University of Southern Denmark\\Odense, Denmark\\pica@cp3-origins.net}

\and
\IEEEauthorblockN{Luigi Del Debbio}
\IEEEauthorblockA{The Higgs Centre for Theoretical Physics\\University of Edinburgh\\Edinburgh, UK\\luigi.del.debbio@ed.ac.uk}
\\
\IEEEauthorblockN{Agostino Patella}
\IEEEauthorblockA{PH-TH\\CERN\\Geneva, Switzerland\\agostino.patella@cern.ch}
\\
\IEEEauthorblockN{Antonio Rago}
\IEEEauthorblockA{School of Computing and Mathematics, \\Plymouth University\\Plymouth, UK\\antonio.rago@plymouth.ac.uk}}


\maketitle
\copyrightnotice

\begin{abstract}
Lattice Quantum ChromoDynamics (QCD), and by extension its parent field, Lattice Gauge Theory (LGT), make up a significant fraction of supercomputing cycles worldwide. As such, it would be irresponsible not to evaluate machines' suitability for such applications. To this end, a benchmark has been developed to assess the performance of LGT applications on modern HPC platforms. Distinct from previous QCD-based benchmarks, this allows probing the behaviour of a variety of theories, which allows varying the ratio of demands between on-node computations and inter-node communications. The results of testing this benchmark on various recent HPC platforms are presented, and directions for future development are discussed.
\end{abstract}


%
\IEEEpeerreviewmaketitle

\section{Introduction}
Quantum ChromoDynamics (QCD), the theory of the strong interaction of quarks and gluons, is a highly successful theory with high-precision predictive power. However, calculations of physical interest are rarely analytically tractable, instead requiring Monte Carlo simulation of a discretised treatment referred to as Lattice QCD (LQCD). Lattice QCD codes are developed by a number of theoretical particle physics research groups internationally, and these codes use a significant fraction of available supercomputing capacity worldwide---for example, NVIDIA quote that up to 20\% of North American supercomputing cycles are used for QCD research \cite{clark}. 

QCD lies in a family of models known as gauge theories, and the numerical techniques developed to study QCD can also be applied to other gauge theories, forming a broader area of research known as Lattice Gauge Theory (LGT). Such theories may differ from QCD in a number of ways; computationally, the difference is typically the dimensionality and structure of the sub-matrices related to each point in the discretised space. These differences can have an impact on the demands that LGT code makes of the computer on which it runs---for example, altering the ratio of computations to communications demands. 

Non-QCD LGT has become of interest recently as a tool for theoretical physicists to probe physics Beyond the Standard Model (BSM), for example relating to recent discoveries at the Large Hadron Collider at CERN. Recent reviews of such techniques include \cite{DelDebbio:2014ata,Giedt:2015alr}.

Benchmarks have previously been developed out of QCD codes, and many of these benchmarks have been adopted in common benchmark suites used for machine evaluation (for example, the NERSC MILC benchmark developed from the MILC research code \cite{milc-benchmark}). However, the QCD codes used for these benchmarks do not have sufficient flexibility to probe BSM theories of physical interest. Thus in order to characterise the diverse performance demands of BSM LGTs, a novel benchmark is necessary, derived from (or at least approximating) a flexible LGT code.

In this work, we present BSMBench, a benchmark satisfying this criterion, derived from the HiRep research LGT code \cite{DelDebbio:2008zf,DelDebbio:2009fd}. In the remainder of this paper, in section \ref{sec:lgts} we will outline the relevant details of LGT (in particular highlighting differences from QCD), then in section \ref{sec:bsmbench}, we will describe the methodology of the benchmark. In section \ref{sec:results} we will then present some selected results, characterising some recent machines' performance in the tests set up by the benchmark, before concluding and suggesting future directions our work will take.

\section{Lattice Gauge Theories}
\label{sec:lgts}

\subsection{Field content}
The ``lattice'' in Lattice Gauge Theory is a hypercubic array of points (``sites''), forming a discretised space (or spacetime). This space is most frequently four-dimensional, and the length in the three spatial directions is generally made to be the same, giving a total number of sites $L^3 \times T$. Each lattice site has eight nearest neighbours (up and down in each of the four dimensions); the lines joining a point to a nearest neighbour is referred to as a ``link''. The lattice typically has periodic boundary conditions, so the number of links is four times the number of sites (avoiding double counting positive and negative links), with no edge corrections.

\newcommand\Nf{\ensuremath{N_\textnormal{f}}}
A gauge theory will typically have one gauge field (the gluon field of QCD), and $\Nf \ge 1$ ``flavours'' of fermion (the quark fields). On the lattice, the gauge field is an $N\times N$ complex-valued matrix on every link, while the fermion fields are $M$-dimensional complex-valued vectors at every site. $N$ and $M$ are integer-valued tunable parameters of the theory; $N\ge 2$ may be freely chosen, while $M \ge N$ is constrained to certain values allowed by group theory (specifically, the $M$-vector must transform under some non-trivial representation of the group of which the $N\times N$ matrix is an element).

Counting up, at each site the gauge field contributes 4 links $\times (N\times N)$ elements $\times$ 2 real numbers; i.e.~$8N^2$ real numbers per site. The fermion fields, meanwhile, contributes $\Nf\times M \times 2 = 2M\Nf$ real numbers per site. The contributions to the whole lattice are then multiplied by $L^3 T$; i.e.~ the gauge field comprises $8N^2 L^3 T$ real numbers, and the fermion fields $2M\Nf L^3 T$. In the case of QCD, $N=M=3$ (the three colors of QCD---red, green and blue).

\subsection{Dirac operator}

The physics of the fermions is encoded in the so-called ``Dirac operator''; on the lattice this is a matrix relating all elements of the fermion field to all elements of the fermion field---that is to say, it is a $(2M\Nf L^3 T)\times(2M\Nf L^3 T)$-element matrix. The interactions in the fermion fields are taken to be nearest-neighbour, resulting in the Dirac operator being exceedingly sparse, and depend on the values of the gauge field elements. The primary task of the Monte Carlo code is to invert this matrix. This is typically done using a Conjugate Gradient (CG) or related algorithm, and as such the dominant (and to a reasonable approximation, only) contribution to the runtime comes from the routine to multiply the fermion field by the Dirac operator; we will call this Dphi\footnote{Some other works refer to this as Dslash.}.

This overwhelming dominance of execution time by one single subroutine has naturally led to it being the focus for optimization; in QCD applications it is not uncommon for Dphi to be hand-optimized with for example vector intrinsics and manual prefetching, rather than written in a na\"ive way and relying solely on the compiler. For more general LGT tools this kind of optimization is less practical; the need for generality in $N$ and $M$ precludes us from hard-coding highly-optimised code in the way that QCD codes can for a fixed $N$ and $M$. For example, the HiRep research code uses a code generator to produce sets of macros for the matrix-matrix and matrix-vector operations required, which are then called from Dphi.

\subsection{Parallelisation}
The hypercubic geometry and nearest-neighbour interactions found in the problem means that it naturally lends itself to spatial parallelisation, with the lattice being sliced up in each dimension, and each resulting piece of lattice being handled by a dedicated process (with processes generally communicating via MPI, although hybrid OpenMP+MPI approaches exist). The need to store and communicate boundary terms places a lower bound on the piece size that can be efficiently handled, and thus an upper bound on the degree of parallelisation for a particular problem size.

\section{The Benchmark}
\label{sec:bsmbench}
The priorities when developing BSMBench were to reflect the computational demands and portability of the HiRep research code, to be able to probe more than one theory (i.e.~set of values of $(N,M)$ above---since the performance demands change as a function of these parameters), to run in reasonable time, and to spend sufficiently long that the run time is not dominated by startup overheads. Additionally, the test suite should be easily run by non-LGT specialists, so that it may, for example, be used by hardware vendors to quote performance of development machines early on in procurement cycles, without having to grant system access to end-users.

The strategy chosen to meet these criteria was based on that of L\"uscher \cite{Luscher:2001tx}. It takes three tasks---two more elementary vector and matrix-vector operations, followed by the full Dphi---and in turn iterates them on randomly-generated fields for a fixed period of time\footnote{To avoid spending excessive time on time checks, the numbers of iterations between time checks doubles after each check.}. (The CG inversion is currently not benchmarked, but can be requested as a check on the machine's numerics.) The number of floating point operations for each task has previously been calibrated, and thus the performance can be quantified by FLOPs/s = Number of iterations $\times$ FLOPs per iteration / Time Taken.

The problem size is fixed, thus the benchmark probes strong scaling behaviour. Since the problem size in production is typically fixed by physical demands, research use of the benchmark is less interested in weak scaling; however, it is possible that it will be added in a future version.

The benchmark is provided with case scenarios, corresponding with three theories: $(N,M)=(2,4)$ (communications-dominated), $(3,3)$ (balanced, and equivalent to QCD), and $(6,6)$ (compute-dominated). Rather than including the full code generator, output header files for these theories are included with the benchmark.

Even for the communications-dominated theory, each iteration has a fixed number of FLOPs, and so the FLOP/s rate for the benchmark gives a proxy to the performance. The advantage of using the same measure for all three theories is that the benchmark statistics may then be directly compared between theories.

BSMBench was constructed by paring down the HiRep research code to the essential elements, thus the benchmarked code closely reflects the workloads in production runs. Further, optimisations can cross-pollinate between HiRep and BSMBench. 

Owing to the need to be able to adopt new HPC infrastructure as it becomes available, the HiRep research code is highly portable---in general, it can be run on a new machine simply by setting the correct compiler and running \texttt{make}. This property is inherited by BSMBench; in the results shown below, no code changes needed to be made to allow the benchmark to run, and to be reflective of the de factor usage of the research code the only optimisation performed was some tuning of the compiler flags. As mentioned, were system vendors to adopt the benchmark and optimise it more aggressively, the optimisations could be backported to the research code to benefit all users.

The flexibility of the benchmark thus manifests in three ways: it is easily portable to many architectures, it can run in reasonable time on a diverse range of machine sizes, and most importantly, it can tune the relative demands on computation and inter-process communication.

\section{Results}
\label{sec:results}
BSMBench has been tested on a variety of HPC platforms, including IBM Blue Gene/P and /Q machines, an SGI ICE XA system with Haswell CPUs, Fujitsu x86 clusters (Westmere and Sandy Bridge-based, at HPC Wales), a Xeon Phi-based cluster (at the Hartree Centre), commodity clusters (both Infiniband and gigabit Ethernet setups), and a Mac Pro workstation.\footnote{The benchmark has also more recently been used on other recent prototype architectures; however, the results of these analyses are currently subject to non-disclosure agreements.} Details of MPI libraries, compilers, and compiler flags are shown in Table \ref{tab:details}; in all cases, the default MPI configuration was used, with no hand-tuning of process placement or run-time flags.

\begin{table*}[!t]
\renewcommand{\arraystretch}{1.3}
\caption{MPI Libraries, Compilers, and Compiler Flags Used to Test Each Machine.}
\label{tab:details}
\centering
\begin{tabular}{|c|c|c|c|}
\hline
\bfseries Machine & \bfseries Compiler & \bfseries MPI & \bfseries Compiler flags \\
\hline
Blue Gene/P & \multirow{2}{*}{IBM XL} & \multirow{2}{*}{IBM (MPICH2-based)} & -O5 -qstrict -qarch=450d -qtune=450 -qunroll -qinline -qhot=simd \\
Blue Gene/Q & & & -O5 -qstrict=precision -qarch=qp -qtune=qp -qhot=level=2 -qsimd\\
\hline
Ethernet cluster & GCC 4.1.2 & OpenMPI 1.3.3 & \multirow{2}{*}{-Wall -std=c99 -O2 -fomit-frame-pointer -mfpmath=sse -msse -msse2} \\
Mac Pro & LLVM-GCC 4.2.1 & MPICH2 1.5 & \\
\hline
Westmere & GCC 4.1.2 & OpenMPI 1.5.4 & -Wall -std=c99 -O2 -fomit-frame-pointer -mfpmath=sse -msse -msse2 \\
Sandy Bridge & Intel 13.0 & Intel 4.1 & -Wall -std=c99 -O3 -xAVX -simd -ipo -finline-functions \\
BlueIce2 & GCC 4.4.6 & OpenMPI 1.6.4 & -Wall -std=c99 -O3 -fomit-frame-pointer -mfpmath=sse -march=native \\
\hline
Xeon Host & Intel 15.0.2 & Intel 5.1.1 & -Wall -ipo -std=c99 -parallel -O3 -xHost \\
Xeon Phi & Intel 15.0.2 & Intel 5.1.1 & -Wall -O3 -ansi-alias -qopenmp -std=gnu99 -mmic\\
\hline
SGI ICE XA & Intel 15.0.5 & SGI MPT 2.14 & -Wall -std=c99 -O2 -xCORE-AVX2 -simd -finline-functions -no-ipo \\
\hline
\end{tabular}
\end{table*}

Full results of each sub-test on every machine tested would be cumbersome to present here, thus we have chosen an interesting subset of results to highlight. Since the Dphi test is most representative of a typical production workload, it is this test that we focus on in presenting results. Results are plotted on a logarithmic scale, to avoid one or two data dominating the plots; plots are shown both of the total FLOP/s, and also of the FLOP/s normalised by the number of processes. In the case of perfect scaling, the latter plot would be a flat line.

\subsection{CPU-based machines}

\begin{figure*}[!t]
\includegraphics[width=\textwidth]{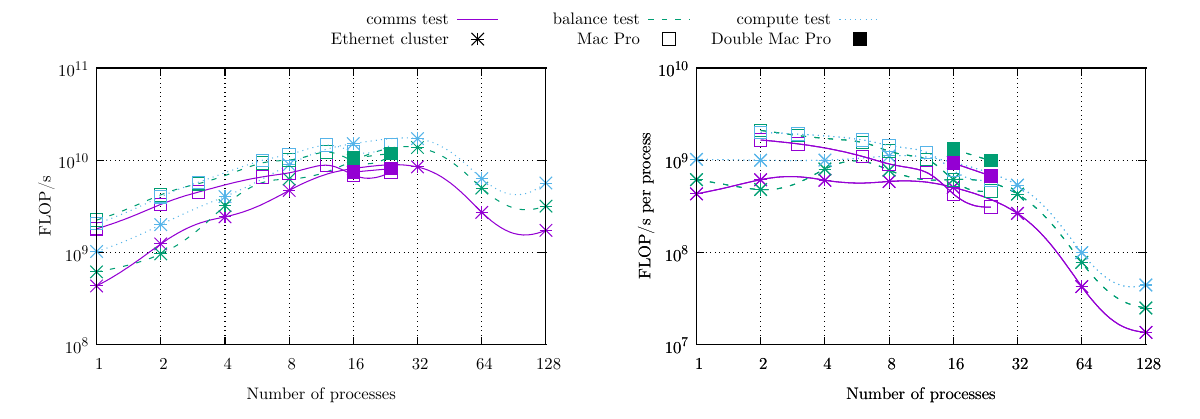}

\vspace{-6pt}\caption{Results for the Dphi test on an Ethernet-only cluster, and a Mac Pro, including a test of HyperThreading performance.}
\label{fig:ulgqcdetc}
\end{figure*}

On machines with all but the most memory-constrained nodes, all tests may be run on a single core, allowing an accurate look at the scalability of the code. We observe this in Fig.~\ref{fig:ulgqcdetc}, where the three tests start off approximately comparably in performance, but the differing communications demands of the three theories used causes the scaling behaviour to differ. The importance of good interconnects for code of this type is clearly demonstrated by the sharp drop-off in performance once the parallelisation goes beyond a single node (16 cores) and starts hitting the network. Also shown are results for a 12-core Mac Pro workstation; this outperforms the cluster on small core counts, but is outperformed core-for-core once core counts increase. We do not expect HyperThreading to give us any advantage on these workloads, since the code makes heavy use of floating-point units, which are shared between the hardware threads.

\begin{figure*}[!t]
\includegraphics[width=\textwidth]{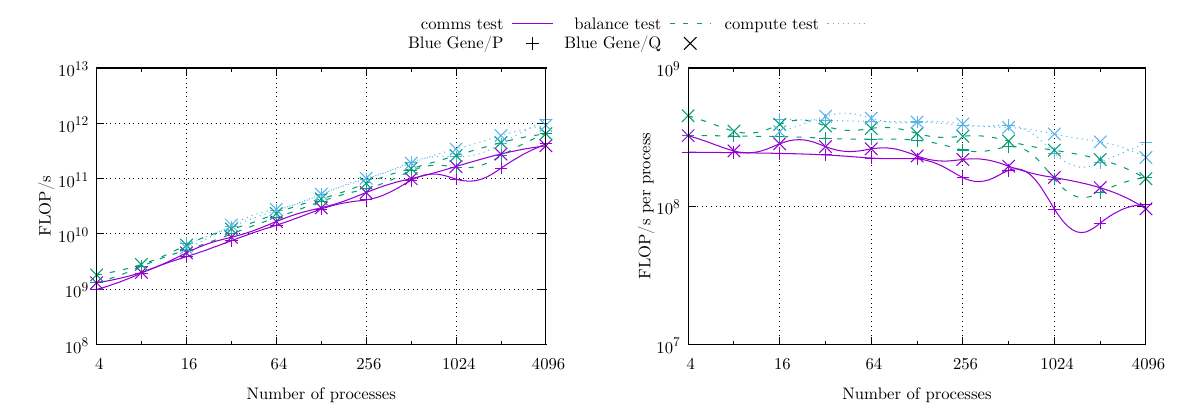}

\vspace{-6pt}\caption{Results for the Dphi test on Blue Gene/P and /Q machines.}
\label{fig:bg}
\end{figure*}

In Fig.~\ref{fig:bg}, both Blue Gene/P and /Q machines show good scaling behaviour; however, core-for-core, the two machines have very similar performance, despite Blue Gene/Q's higher clock speed. The reason for this is vectorisation; as mentioned above, the code does not vectorise well to vectors longer than 2 double-precision floating-point numbers. This means that the 2-double vector units on Blue Gene/P can be used, but not the 4-double vector units on Blue Gene/Q. The performance on Blue Gene/Q fluctuates more as a function of number of processes than on Blue Gene/P; this illustrates the need to tune the process placement to take advantage of the network topology---in the case of Blue Gene/P, the problem sits advantageously on the network topology without the need for optimisation, whereas on Blue Gene/Q, the default layout is non-optimal for some parallelisations.

\begin{figure*}[!t]
\includegraphics[width=\textwidth]{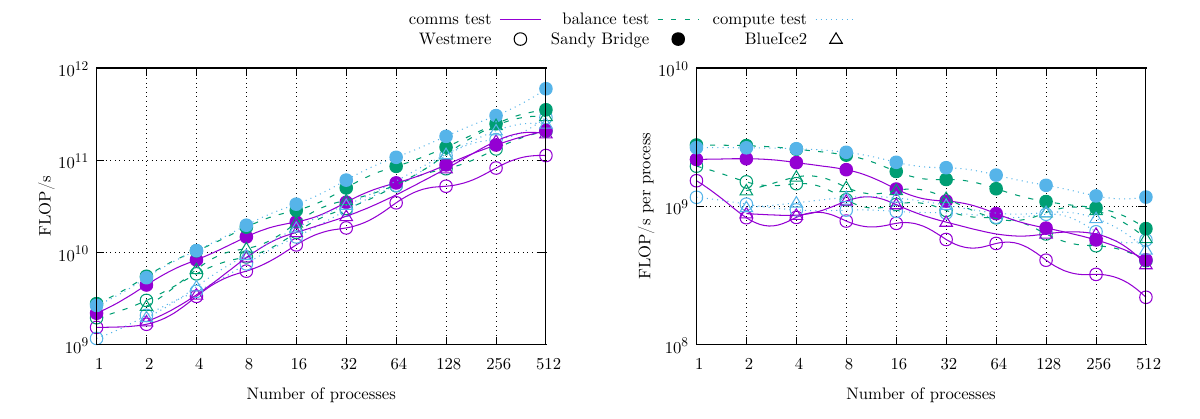}

\vspace{-6pt}\caption{Results for the Dphi test on Intel Westmere- and Sandy Bridge-based HPC Wales clusters, and a Westmere-based cluster (BlueIce2) in Swansea.}
\label{fig:hpcw}
\end{figure*}

Fig~\ref{fig:hpcw} shows results for three clusters; two HPC Wales clusters, one Intel Westmere-based and one Sandy Bridge, and one other Westmere-based cluster (BlueIce2 at Swansea). As we might expect, at low process count the two Westmere clusters perform very similarly, while the Sandy Bridge cluster offers modest ($\sim2\times$) improvements in performance. At higher process counts, the two Westmere systems diverge somewhat; this is due to a greater freedom in choosing the job layout on this system, with 8 rather than 12 MPI tasks per processor dividing more nicely into the powers of 2 in the spatial parallelisation. Also as expected, the performance at low process count is very close between the three theories, but starts to diverge once communications starts to play more of a role.

\begin{figure*}[!t]
\includegraphics[width=\textwidth]{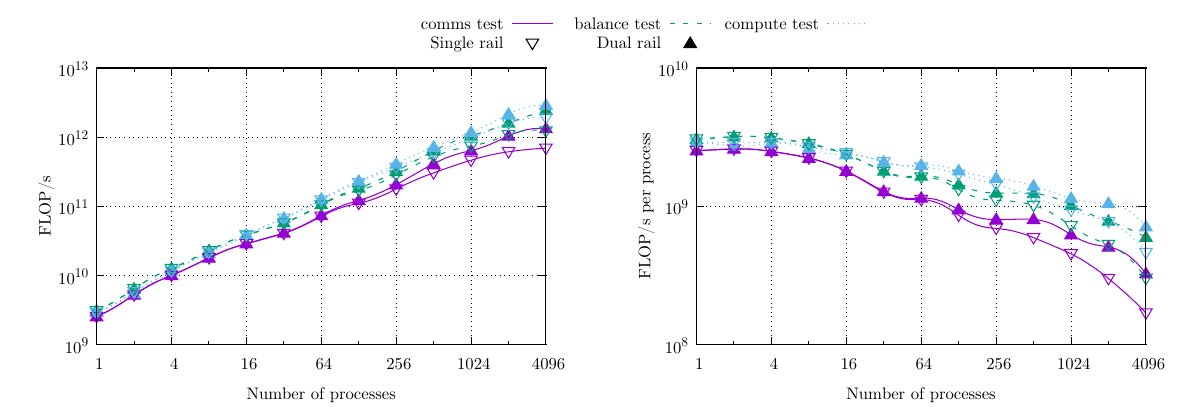}

\vspace{-6pt}\caption{Results for the Dphi test on an SGI ICE XA system with Haswell CPUs.}
\label{fig:sgi}
\end{figure*}

In Fig. \ref{fig:sgi} we show results for an SGI ICE XA system with Intel Haswell nodes (24 cores per node). This system has a dual-plane enhanced hypercube interconnect topology (i.e. there are two independent interconnect fabrics, each with their own switches and cables). The figure shows the benchmark results for both the single-plane (one fabric only) and dual-plane (both fabrics work cooperatively) cases. At low process counts, the per-core performance is similar to the previous-generation Intel architectures; however, at higher parallelisations, the system demonstrates significantly better scaling. The effect of increasing the inter-node communications bandwidth is particularly visible in the comms test.

To briefly summarise these results, all machines tested that have high-speed interconnects (i.e.~not Ethernet) show very good strong scaling at small to intermediate parallelisations. Those with more advanced interconnects (Blue Gene) show better strong scaling at the highest parallelisations than machines with simpler Infiniband arrangements. Both of these effects are most pronounced for the more communications-intensive task. All of the larger machines tested were able to reach between $10^{11}$ and $10^{12}$ FLOP/s. Blue Gene required 4--8 times as many cores to reach comparable performance to x86.

Does this mean, then, that any machine with a fast interconnect is suitable or preferable? This depends on a number of factors. At smaller problem sizes, the maximum parallelisation is reached more quickly, and so it would be preferable to minimise the use of the communications links (by using only one or a small number of nodes) rather than needing to procure the fastest available links. 
Larger problem sizes are only tractable through parallelisation, so the need for the fastest available interconnects becomes more pronounced. This analysis places Blue Gene/P and /Q as similarly desirable on a per-core basis; however, other obvious considerations then come into play---for example, that the footprint and power demands of a Blue Gene/Q would be significantly lower than those of an equivalent number of Blue Gene/P cores.

\subsection{Xeon Phi}
\begin{figure}[!t]
\centering
\includegraphics[width=\columnwidth]{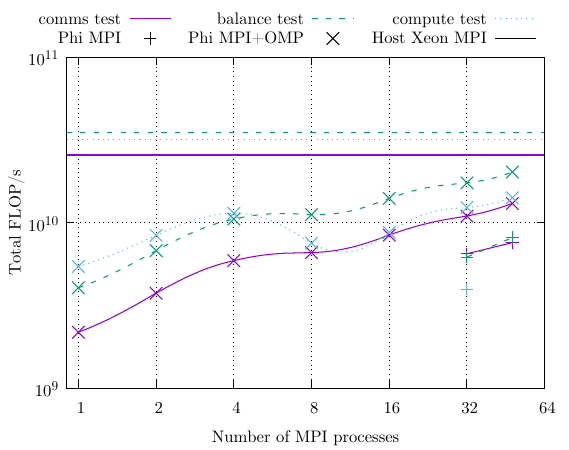}

\caption{Results of testing the performance of a Xeon Phi node at the Hartree Centre. The horizontal lines are the performance of the host node, with two Xeon sockets, using 24 MPI processes (Per-process performance is not shown, since for the OpenMP runs values for the number of processes should attempt to use the entire card, with only the breakdown between MPI and OpenMP changing.)}
\label{fig:phi}
\end{figure}

The benchmark has also been tested on a Xeon Phi (Knights Corner) system at the Hartree Centre. No modifications were necessary to allow the code to compile (beyond specifying the compiler). While the Phi needs 240 threads to keep all cores occupied and gain maximum performance, it was impossible to run that many MPI tasks due to the size of the required MPI buffers exceeding the card's memory. It was therefore necessary to use a hybrid MPI+OpenMP approach. Fig.~\ref{fig:phi} shows the results of these tests; where OpenMP was used, the number of threads was chosen as $\textnormal{Number of threads} = \left\lfloor \frac{240}{\textnormal{Number of MPI tasks}} \right\rfloor$. For the comms and balance test, clearly the hybrid approach gives a performance gain over straight MPI; however, maximising the number of MPI tasks also improves performance over using OpenMP only. (The drop in performance of the compute test between 4 and 8 MPI tasks is currently poorly understood.) The performance is at best approximately half that of the two Xeon sockets on the host; one would hope that this relationship could be inverted if the code could be adapted to make use of the 512-bit vector unit in the KNC processor.

\section{Conclusion}
We have developed a novel benchmark, BSMBench, based on Beyond the Standard Model Lattice Gauge Theory. Unlike previous benchmarks based on QCD, it has the capacity to adjust the theory under study, and consequently modify the workload's demands in terms of the ratio of computations to communications. Thanks to this, BSMBench could be applied in a variety of user scenarios (e.g.~as a monitoring and fault diagnostic tool or as a general-purpose performance evaluation utility) that transcend its original goals.
\newpage
We have tested this benchmark on a variety of recent supercomputing platforms, including CPUs and Xeon Phi coprocessors. Our results show good strong scaling in the presence of a sufficiently fast interconnect, and exhibit the expected splitting between theories under study.

One limitation of the benchmark (and the underlying research code) is an inability to make use of vector units wider than two double precision floating-point numbers; work is underway to lift this restriction, which would significantly boost the performance on more modern architectures featuring AVX and QPX vector instructions. Other future improvements to BSMBench will be to reduce the reliance on parameter sets, with the core code instead able to calculate the necessary parameters, and potentially to introduce a weak scaling test.

Interesting potential tests of the benchmark currently under investigation include assessing the relative performance of different MPI libraries on the same architecture, and observing how the results compare with those of other benchmarks---for example, those based on QCD (for example, \cite{milc-benchmark}), and those based on Conjugate Gradient solvers (for example, \cite{Dongarra01032016}).

BSMBench is available at \url{http://www.bsmbench.org/}.

\section*{Acknowledgment}
The authors would like to thank DiRAC, the Hartree Centre, HPC Wales, IBM, SGI, and the University of Liverpool for allowing access to machines for benchmarking purposes. 
Additional computations were performed on machines in Swansea University. EB would like to thank IBM Research for their hospitality during the early stages of this work.




\bibliographystyle{IEEEtran}
\bibliography{references}
%

\end{document}